# LAMP: A Locally Adapting Matching Pursuit Framework for Group Sparse Signatures in Ultra-Wide Band Radar Imaging

Sanghamitra Dutta and Arijit De

*Abstract*— **It has been found that radar returns of extended targets are not only sparse but also exhibit a tendency to cluster into randomly located, variable sized groups. However, the standard techniques of Compressive Sensing as applied in radar imaging hardly considers the clustering tendency into account while reconstructing the image from the compressed measurements. If the group sparsity is taken into account, it is intuitive that one might obtain better results both in terms of accuracy and time complexity as compared to the conventional recovery techniques like Orthogonal Matching Pursuit (OMP). In order to remedy this, techniques like Block OMP have been used in the existing literature. An alternate approach is via reconstructing the signal by transforming into the Hough Transform Domain where they become point-wise sparse. However, these techniques essentially assume specific size and structure of the groups and are not always effective if the exact characteristics of the groups are not known, prior to reconstruction. In this manuscript, a novel framework that we call locally adapting matching pursuit (LAMP) have been proposed for efficient reconstruction of group sparse signals from compressed measurements without assuming any specific size, location, or structure of the groups. The recovery guarantee of the LAMP and its superiority compared to the existing algorithms has been established with respect to accuracy, time complexity and flexibility in group size. LAMP has been successfully used on a real-world, experimental data set.**

## I. Introduction

ULTRA WIDE BAND (UWB) technology [1] though still in its phase of development, finds versatile use in various civilian and defense applications. The recent applications of impulse radars in health-care sector [2] for remote monitoring of physiological parameters and medical imaging have led to its widespread popularity. The potential of impulse radio [3] exploiting nanosecond pulses for short range communications in dense multipath environments is also being explored in recent years. In defense and military applications, UWB impulse Radar finds special importance in extending the human vision beyond mediums through which visible light cannot penetrate. This principle has been exploited in several applications like ground penetrating radar (GPR) [4] and through-wall radar imaging (TWRI) [5]. Impulse radar for such applications utilizes short duration pulses [6] having a bandwidth of few GHz, typically ranging from 300 MHz on the lower end to 6 GHz in the upper end of the spectrum. Higher frequencies increases the resolution of imaging, but suffers from attenuation and decrease in depth of penetration. Impulse radar has many advantages over the conventional radar systems, providing higher range measurement accuracy and better range resolution, improved immunity to interferences and multipath, better penetrating capability, better object detection probability and reliable object tracking.

UWB impulse radar generates and transmits short pulses (usually Gaussian or its derivative) through a transmitting antenna (TX) that propagates through a lossy dielectric medium. When the pulses hit an object in their path, a part of the electromagnetic energy is scattered from the object and propagates back to receiving antenna (RX), which is usually located at the same position as the transmitter antenna(TX) for a mono-static imaging system. The time (range) profile between the transmitted and received signal corresponds to the round trip spatial distance between TX/RX and the visually hidden targets, resulting in what is commonly referred to as a single A-scan consisting of N time samples. To identify the target, several measurements consisting of a set of P antennae positions placed along a single trajectory, each collecting several A-scan data, results in a 2D image, commonly referred to as a B-scan measurement as depicted in Fig.1.

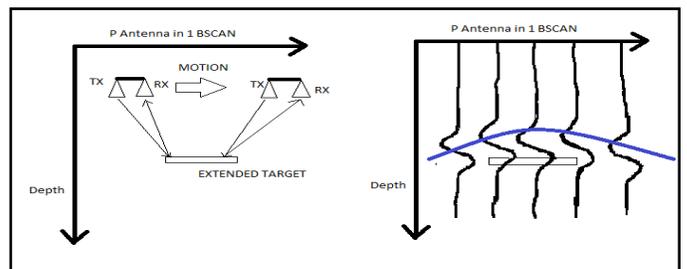

Fig 1. Schematic of a single B-scan imaging of an extended buried target

High resolution imaging demands acquisition and processing of large amount of data samples at high sampling rate. In order

Sanghamitra Dutta (email: sanghamitra2612@ece.iitkgp.ernet.in) and Arijit De (email: arijit@ece.iitkgp.ernet.in) are with the Department of Electronics and Electrical Communication Engineering at the Indian Institute of Technology-Kharagpur, Kharagpur – 721302, India.

to remedy this, sub-Nyquist sampling techniques for fast data acquisition were introduced via the "Compressed Sensing' paradigm in GPR [7,8] and TWRI applications [5].

The main idea of Compressive Sensing in radar imaging is that instead of storing the traditional samples for every antennae location, a small set of informative measurements in the form of randomized projections as shown in Fig.2 is sufficient to reconstruct the image of the buried target.

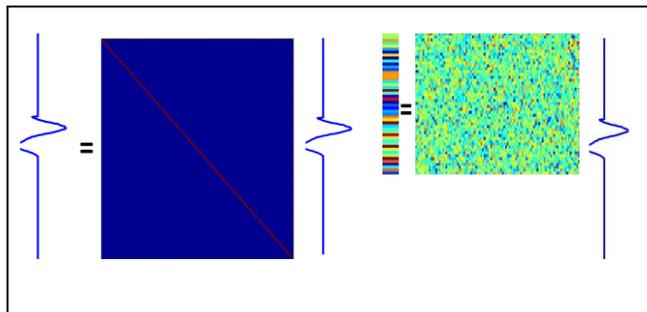

Fig. 2: Schematic of (a) Uniform Sampling (left) and (b) Compressive Sampling (right). The elements are normalized within the range 1 (red) and 0 (blue).

Typically two approaches have been considered in recent literature: stepped-frequency approach and time-domain approach. In the former case as in [7], an operator was defined to map the target-space into the frequency domain and random frequency domain samples were taken at each antenna locations. The target locations, being sparse, this methodology enabled efficient localization with less number of frequency samples than would have normally been required for traditional stepped-frequency GPR systems. However construction of such operator involved additional knowledge of the propagating medium and the assumption that the RCS of the target, soil permittivity etc. was constant with respect to the wideband frequency range of such imaging systems. A different approach has been discussed in [8], where the sparse target-space was mapped to selected dictionary of the known transmitted pulse and a small set of randomized projections of the time-domain samples were considered to estimate the target locations. However, construction of such a dictionary again involves knowledge of the nature of the reflected signal, signal spread function and the response of the target. In the scenario of extended targets, the scattered signal is not necessarily the same as the incident pulse and comprises of an early time specular reflection and a late-time ringing which characterizes the creeping wave around the target [9]. Most of the existing literatures on Compressive Sensing in TWRI have also considered such a mapping operator or dictionary from target space to frequency or time domain [5] and hence have its inherent limitations. In this manuscript however, instead of mapping the space of point targets onto frequency or time domain, we attempt to reconstruct the original time domain scattered signal itself from fewer measured signal data, by exploiting the property that the reflected signals in the time-domain are clustered around the target locations. This method gives a more reliable and accurate estimation compared to the case when the additional knowledge about the propagating medium or nature of reflected signals are not known and hence construction of a proper operator or dictionary as mentioned in the aforementioned techniques are not feasible.

A possible hardware implementation of compressive signal acquisition scheme for time domain signals has been discussed in [8]. The authors have considered different types of random sensing matrices with elements drawn from standard Normal distribution $N(0,1)$ or Bernoulli distribution i.e. $\pm 1$ with equal probability, which can be implemented in hardware by some of the recently proposed architectures such as the Random Demodulator (RM) [11], Modulated Wideband Converter [12] and Random Modulator Pre Integrator (RMPI) [13]. Readers are referred to [14] for a detailed review. The RMPI, as depicted in Fig. 3 usually modulates the analog input signal with random sequences in parallel channels, integrates (a low pass filtering operation) them and then samples the output from each channel using low frequency ADCs. Though the mixing in each channel is done at the Nyquist frequency, its main advantage lies in the fact that mixing a signal at high frequency is easier compared to sampling it accurately at that frequency. For special choice of binary sensing matrices as in [10], it essentially boils down to switching on or off the low noise amplifier at the input of the receiver. Thus using a proper design of RF mixers and filters one can acquire compressed measurements using low frequency ADCs, instead of sampling the entire signal at a higher sampling rate, leaving the tedium of efficient recovery of the original time domain samples on the processor.

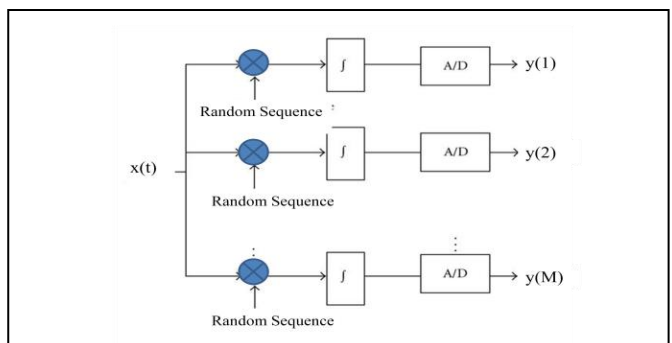

Fig. 3: Typical RMPI Architecture for radar Imaging [14]

Various algorithms [16-21] have been explored in literature for the reconstruction of a sparse signal from compressed measurements which can be broadly classified into convex optimization based approach, greedy approach and Bayesian with prior information. Greedy algorithms provide fast and simpler implementation compared to other techniques [20]. Rarely have the clustering tendency of the time domain scattered signals from the target been considered in the process of reconstruction. In this manuscript we provide with proper justifications a locally Adapting greedy-type algorithm, which we refer to as Locally Adapting Matching Pursuit (LAMP), an advancement of the traditional and quite popular Orthogonal Matching Pursuit (OMP) [16, 19] based greedy approach. This

14$^{th}$ November, 2014 DRAFT

generic approach can be extended to any greedy algorithms for efficient recovery of time domain radar signals from its compressed information.

The rest of the paper is organized as follows. The basic problem formulation and motivation for LAMP is discussed in Section II. Section III provides a review of the existing signal recovery techniques. Section IV describes the proposed algorithm for a single A-Scan measurement vector establishing the recovery guarantee of the LAMP algorithm and compares the results obtained using LAMP with the existing techniques. Various implementation aspects in improving the performance of the algorithm are discussed in this section. Section V extends the algorithm to compressively acquired B-Scan data. We conclude the manuscript with possible future improvements of the proposed algorithm in Section VI.

## II. SPARSE PROBLEM PRELIMINARIES

Once the reduced dimension radar data has been acquired by compressed sampling as discussed before, the problem now boils down to reconstruction of the original time domain samples of larger dimension than the acquired data. For a single A-scan, the measurement model can thus be represented as,

$$\mathbf{y} = \mathbf{A}\mathbf{x} \quad (1)$$

where,

(i) $\mathbf{x}$ is the actual time-domain signal vector of dimension $N \times 1$ consisting of N samples in time domain

(ii) $\mathbf{A}$ is the $M \times N$ dimensional sensing matrix used in data acquisition with $M \ll N$ as shown in Fig. 3, with $\mathbf{A} = [\mathbf{A}_1 \mathbf{A}_2 .... \mathbf{A}_N]$

(iii) $\mathbf{y}$ is the acquired compressed reduced set measurements of dimension $M \times 1$.

The above representation is defined as Single Measurement Vector (SMV) model. Clearly, as $M \ll N$, there exists many solutions to the underdetermined system of equations stated in (1). In order to find a unique solution to the aforementioned problem, additional constraints are generally imposed on the signal vector $\mathbf{x}$. It might be observed that, typically in radar imaging the signal vector $\mathbf{x}$ consists of very few non-zero components in the time domain and hence $\mathbf{x}$ is referred to as a sparse vector. The number of non-zero elements in the vector can be expressed as $|\mathbf{x}|_0^0 \leq K$ with $K \ll N$. Such a signal is often called a K-sparse signal. The objective of sparse recovery algorithms is therefore, to find the sparse vector $\mathbf{x}$, given $\mathbf{A}$ and $\mathbf{y}$. The problem can be represented as:

$$\min |\mathbf{x}|_0^0$$
$$\text{Such that } \mathbf{y} = \mathbf{A}\mathbf{x} \quad (2)$$

Problem (2) is NP-hard and combinatorial in nature, hence cannot be solved in polynomial time. In an elegant paper [12], Candes and Tao have demonstrated that under certain restrictions on the sensing matrix A, the solution to the problem (2) is unique and equivalent to the relaxed convex problem (3) stated below as,

$$\min |\mathbf{x}|_1$$
$$\text{Such that } \mathbf{y} = \mathbf{A}\mathbf{x} \quad (3)$$

The problem (3) which involves minimizing the $l_1$-norm can be solved using convex optimization techniques or greedy pursuits [16-21], the details of which will be discussed in the following section. In most cases, Orthogonal Matching Pursuit (OMP) [16, 19] has been used for image reconstruction. The technique has been found to be effective if the signal vector $\mathbf{x}$ is sparse in the spatial domain. In many practical applications, additional structure might be incorporated to correctly reflect the property of the signal. A typical example is when the K non-zero coefficients are grouped together into randomly located variable sized clusters as shown in Fig. 4 below. For such signals typically referred to as Group-sparse, OMP has not been found to be satisfactory. The group sparsity are not always correctly captured using the conventional OMP since the group sparse solution might not always be the sparsest solution, but might arise in typical scenarios like the case of time domain signal in the case of radar imaging.

(a)

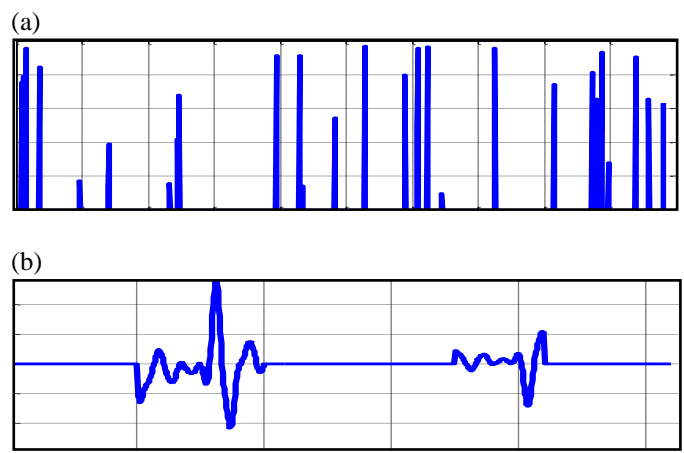

(b)

Fig 4. (a) A general sparse signal (Top) and (b) A group sparse signal (Below) typically encountered in radar imaging.

Group Sparsity has also been studied incorporating some modifications in dictionary learning or using Block Orthogonal Matching Pursuit (BOMP) in [24-26]. A compressive Bayesian Framework for group sparse signals has been introduced in [27] and the concept was applied to TWRI in [23]. However, these algorithms have either assumed a probabilistic framework as in Bayesian learning or prior knowledge of the group structure and block size as in BOMP. In this manuscript, an improved algorithm has been proposed that is principally different from those stated above, in the sense that it does not assume any possible group locations, nature or prior distribution of the groups. It adaptively determines the size and structure of the group once a seed is located within the group. It can thus be used for identifying variably sized, randomly located groups including some groups which might consist of only a single data point.

In a similar approach as in (2), the B-scan data is traditionally represented by an SMV model by concatenating the A-scan compressed data for each antenna positions into a single vector [5,Ch-14] as,

$$\mathbf{y'} = \mathbf{A'x'} \qquad (4)$$

where,

(i) $\mathbf{x'} = [\mathbf{x}_1^T \mathbf{x}_2^T ....\mathbf{x}_P^T]^T$ is a vector of dimension $NP \times 1$ formed by concatenation of the measurement data obtained from each antennae location. Each column vector $\mathbf{x_p}$ denotes the original $N$ samples of the time-domain measurements corresponding to the $p$-th antennae location for $p = 1, 2, .., P$.

(ii) $\mathbf{A'}$ is the $M' \times NP$ dimensional sensing matrix with $M' \ll NP$

(iii) $\mathbf{y'}$ is a $M' \times 1$ dimensional compressive measurement vector representing the total B-scan data.

It might be noted that, such a representation does not explore the correlation between the received signals at the antennae placed at adjacent positions that arises mainly in the case of extended targets. An approach to detect extended buried object has been addressed by transforming the B-scan image into Hough Transform Domain in [28]. However, when exact knowledge of the nature of the groups in GPR Imaging is not known, the concept of parameterized shapes has not been found to be particularly useful. In this manuscript, an alternate approach has been proposed where the correlation in the received signals in adjacent antennae locations has been explored by placing each A-scan signal vector adjacently in the form of a matrix, that is referred to as the Multiple Measurement Vector (MMV) [29] model. The proposed algorithm which exploits the group sparsity has been extended for the MMV model and is described in Section V.

## III. EXISTING RECOVERY TECHNIQUES

The fundamental problem of sparse reconstruction as shown in (1) deals with recovery of a $K$-sparse vector $\mathbf{x}$ of dimension $N$, from a smaller subset $\mathbf{y}$ consisting of $M$ compressive measurements. In the context of radar imaging, the returned signal from the target $\mathbf{x}$ is generally group sparse signal consisting of combination of Gaussian pulses and its derivatives of varying widths and time delays. The sensing matrix $\mathbf{A}$ as discussed before is typically a random Gaussian or Bernoulli Matrix of dimensions $M \times N$ with $M \ll N$. In the sparse recovery framework, this underdetermined system of linear equations (1) can be solved using various approaches of which Basis Pursuit [17], and LASSO [30] are most popular. Basis Pursuit is an optimization principle, introduced by Chen and Donoho in [17] which directly solves (3) by translating it into a linear programming problem that can be solved using various convex optimization techniques [18]. Alternately, LASSO as proposed by Tibshirani in [30] attempts to solve the following optimization problem with an added penalty function as,

$$\min_{\mathbf{x}} |\mathbf{y} - \mathbf{Ax}|_2^2 + \lambda |\mathbf{x}|_1 \qquad (5)$$

where $\lambda$ is a regularizing parameter trading the smoothness of the approximation with the level of sparsity. Choice of $\lambda$ is crucial for correct recovery the desire signal.

A second type of approach typically used in solving the above problem includes Greedy Algorithms for selection of a subset of $\mathbf{x}$ which has non-zero components. Popular among them are Orthogonal Matching Pursuit (OMP) [16, 19], Compressive Sampling Matching Pursuit (CoSaMP) [21] Iterative Thresholding [33] and their various extensions. For a detailed discussion on the various existing recovery techniques, the reader is referred to [32]. The major advantage of greedy algorithms over convex optimization based techniques is its speed, simplicity, and feasibility for hardware implementation [31]. In this paper we have extended the OMP framework to account for the property of group-sparsity for signals which arises in the context of UWB radar scattering. Our proposed approach will be discussed in the following section and can be extended to any other greedy algorithms mentioned above. The technique of Orthogonal Matching Pursuit (OMP) and its related counterpart: Orthogonal Least Squares (OLS) which has also been in vogue in sparse recovery will be discussed henceforth.

Before proceeding further, the notations used here are briefly discussed. Suppose the index set $T \subseteq \{1, 2...N\}$, $\mathbf{x}|_T$ denotes a vector of length $N$ such that $\mathbf{x}|_T(j) = \mathbf{x}(j) \quad \forall \ j \in T$ and zero otherwise. Similarly, $T^c$ is defined as the set $\{1, 2, ..., N\}/T$ and $\mathbf{x}|_{T^c}$ denotes a vector of length $N$ such that $\mathbf{x}|_{T^c}(j) = \mathbf{x}(j) \quad \forall \ j \in T^c$ and zero otherwise. By $\mathbf{A}|_T$ we denote the sub-matrix of dimension $M \times |T|$, consisting of the columns of $\mathbf{A}$ as indexed by $T$. Let the projection operator onto the column span of $\mathbf{A}|_T$ be denoted as $\mathbf{P}_T = \mathbf{A}|_T (\mathbf{A}|_T)^\dagger$ where $\mathbf{A}|_T^\dagger = ((\mathbf{A}|_T)^\mathbf{T}(\mathbf{A}|_T))^{-1}(\mathbf{A}|_T)^\mathbf{T}$ denotes the well known Moore Penrose Pseudo Inverse. Similarly, $\mathbf{P}_T^\perp = (\mathbf{I} - \mathbf{P}_T) = (\mathbf{I} - \mathbf{A}|_T (\mathbf{A}|_T)^\dagger)$ denotes the projection operator onto the orthogonal complement of the column span of $\mathbf{A}|_T$. Let $\mathbf{Q}|_T = \mathbf{P}_T^\perp \mathbf{A}$ be the matrix obtained as a result of orthogonalising the columns of $\mathbf{A}$ against the column span of $\mathbf{A}|_T$. We also define $\mathbf{h}_T = (\mathbf{Q}|_T)^T \mathbf{Q}|_T \mathbf{x}|_{T^c}$ and $Supp(\mathbf{x})$ as the support set consisting of indices corresponding to the non-zero components of $\mathbf{x}$. The function $residue$ is defined as $residue(T') = \mathbf{P}_{T'}^\perp \mathbf{y}$ where $T' \subseteq \{1, 2, ...., N\}$ is any index set.

### A. Orthogonal Matching Pursuit (OMP)

OMP [16] is a one of the most widely used greedy algorithm in sparse recovery. It is primarily based on the idea that if the columns of the sensing matrix $\mathbf{A}$ as given in (1) are "nearly orthonormal", then the columns of $\mathbf{A}$ corresponding to the elements in the support of $\mathbf{x}$ will have significant contribution in the linear combination that forms the vector $\mathbf{y}$. Thus, the columns of $\mathbf{A}$ should have higher projection onto the vector $\mathbf{y}$ corresponding to the indices that actually belong to the support of $\mathbf{x}$, which is known to $K$-sparse.

Let us assume that $T$ denotes the set of indices of $Supp(\mathbf{x})$ selected up to the current iteration. We are thus required to find the support of $\mathbf{x}|_{T^c}$. Clearly $T = \phi$ at the beginning of the first iteration. Let us also define the residue obtained after the nth iteration as $\mathbf{r}_n = \mathbf{P}_T^\perp \mathbf{y}$. Note that $\mathbf{r}_0 = \mathbf{y}$ at the beginning of the first iteration. OMP updates the set $T$ using the following selection criteria,

$$j = argmax |<\mathbf{A}_j, \mathbf{r}_n>| \qquad (6)$$

At each iteration, the set $T$ is updated as $T = T \cup \{j\}$ and the new residue is obtained as $\mathbf{r}_n = \mathbf{P}_T^\perp \mathbf{y}$ with the updated $T$. The selection process continues until $K$ elements are found or if the residue is no longer significant.

Note that, from [18] $\mathbf{h}_T$ can be written as,

$$\begin{aligned}\mathbf{h}_T &= (\mathbf{P}_T^\perp \mathbf{A})^T (\mathbf{P}_T^\perp \mathbf{A} \mathbf{x}|_{T^c}) \\ &= (\mathbf{P}_T^\perp \mathbf{A})^T (\mathbf{P}_T^\perp \mathbf{A} \mathbf{x}) \\ &= \mathbf{A}^T \mathbf{P}_T^\perp \mathbf{y} = \mathbf{A}^T \mathbf{r}_n \end{aligned} \qquad (7)$$

using symmetry and idempotence property of the projection operator $\mathbf{P}_T^\perp$. Thus, the selection rule of OMP might also be written as,

$$j = argmax |\mathbf{h}_T(j)| \qquad (8)$$

Tropp and Gilbert [16] have proven that under certain restrictions on the sensing matrix $\mathbf{A}$, OMP provides the correct recovery of the sparse signal $\mathbf{x}$ with a high probability. An alternate analysis of OMP has been provided by Davenport and Wakin [18] using the *Restricted Isometry Property* (RIP). A matrix $\mathbf{A}$ satisfies RIP of order $K+1$ if there exists a constant $\delta \in (0,1)$ such that,

$$(1-\delta)|\mathbf{x}|_2^2 \leq |\mathbf{A}\mathbf{x}|_2^2 \leq (1+\delta)|\mathbf{x}|_2^2 \qquad (9)$$

which holds for all $\mathbf{x}$ with $|\mathbf{x}|_0^0 \leq K+1$. In [18] the authors have provided a recovery guarantee for OMP if the sensing matrix satisfies RIP of order K+1, with isometric constant $\delta < 1/(3\sqrt{K})$ where $K$ is the size of $Supp(\mathbf{x})$. Hence, choice of appropriate sensing matrix is important for exact recovery using OMP.

The idea of OMP has also been extended to Multiple Measurement Vectors (MMV) in [29] and will be discussed in Section V. A selection criterion for updating the choice of subset, alternate to that in (6), which at certain cases provides better results than OMP, is discussed in the following.

### B. *Orthogonal Least Squares (OLS)*

Orthogonal Least Squares (OLS) [35], often confused with OMP is also a greedy sparse recovery technique that is based on the similar idea of OMP, but differs slightly in the selection rule. At each iteration OLS updates the current index set $T$ as $(T \cup \{j\})$ for,

$$\begin{aligned} j &= argmin_j | residue(T \cup \{j\})| \\ &= argmin_j |\mathbf{P}_{T \cup \{j\}}^\perp \mathbf{y}| \end{aligned} \qquad (10)$$

The similarity and differences of OLS and OMP and its recovery guarantees has been discussed elaborately in [35]. It has also been observed that when the number of measurements $M$ is less, OLS performs better than OMP. However, OLS is expensive in terms of time complexity as in each step it involves calculation of least-squares to calculate the residue for all elements.

### C. *Algorithms for Group Sparse Recovery*

For signals exhibiting group sparsity within blocks of known size $d$, the conventional OMP can be made efficient using a greedy block based OMP (BOMP) as in [24]. In this case, the entire set of indices $\{1, 2...N\}$ is divided into blocks of size $d$, which is assumed beforehand. Instead of finding projections for every single column of $\mathbf{A}$ as in (6), a sub-matrix consisting of the columns of $\mathbf{A}$ whose indices correspond to an entire block, is used in calculating the projection as shown in Fig. 5.

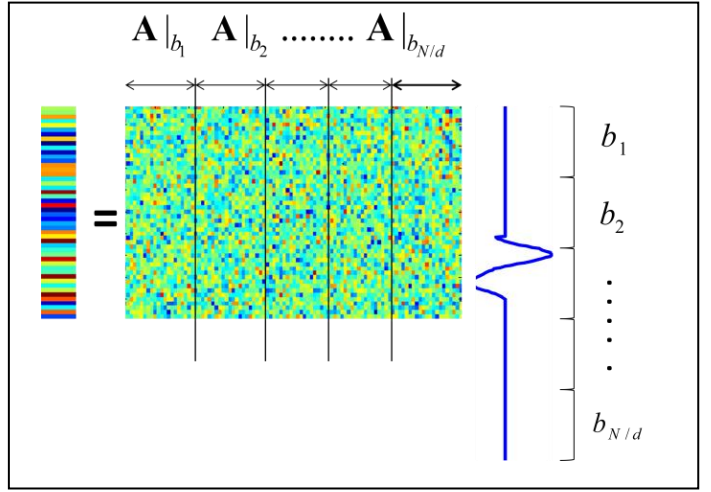

Fig. 5. Representation of the block structure for BOMP and Group Lasso

Let $\mathbf{A}|_{b_1} \mathbf{A}|_{b_2} ........ \mathbf{A}|_{b_{N/d}}$ denote the sub-matrices of size $M$ x $d$ each as shown in the Fig.5. The set $\{1,2,...N\}$ of size $N$ is also divided into $N/d$ disjoint subsets of size $d$, denoted by $b_1$, $b_2$ .... $b_{N/d}$. Note that $b_l = \{(l-1)d+1, (l-1)d+2 ...... ld\}$ where $l$ varies from 1 to $N/d$. At each iteration, the BOMP updates the current support $T$ as $T = T \cup b_l$ where,

$$l = argmax |(\mathbf{A}|_{b_l})^T \mathbf{r}_n|_1 \qquad (11)$$

Here, since $(\mathbf{A}|_{b_l})^T \mathbf{r}_n$ is a vector, we consider the *l*-1 norm of this vector. The new residue is updated as $\mathbf{r}_n = \mathbf{P}_T^\perp \mathbf{y}$ with the updated $T$.

Similar to BOMP, an approach is followed in Convex Optimization that is termed as the Group LASSO [34]. Group

Lasso, which is an extension of standard Lasso (5) attempts to solve the following problem.

$$\min_{\mathbf{x}} |\mathbf{y} - \mathbf{A}\mathbf{x}|_2^2 + \lambda \sum_{l=1}^{N/d} |\mathbf{x}|_{b_l}|_2 \quad (12)$$

Note that, both in BOMP and Group Lasso, the possible block sizes and their locations have to be assumed beforehand, and it only identifies which of those possible blocks are present in the support of $\mathbf{x}$. However, in a practical scenario, if the groups are randomly located and variably sized, it is not possible to assume such possible locations of the groups. This is a primary drawback of these algorithms and leads to the necessity for Locally Adapting Matching Pursuit (LAMP).

In the next section, we describe our proposed algorithm (LAMP) and establish recovery guarantees for the algorithm. A comparison with existing recovery techniques has also been presented.

## IV. LAMP: THE PROPOSED ALGORITHM

LAMP is designed to identify the support of the unknown vector $\mathbf{x}$ with the additional information that it is group sparse. It uses an Adapting searching technique once it identifies a starting point in a group. It initially takes the sensing matrix $\mathbf{A}$ and the compressed measurements $\mathbf{y}$ as input. At every iteration, the current iteration index is stored in variable $n$ and $Supp(\mathbf{x})$, obtained till the current iteration is stored in set $T$. Therefore, the starting conditions are taken as $n = 0, T_0 = \phi$. The residue after the $n^{th}$ iteration is stored in $\mathbf{r}_n$. Initially, $\mathbf{r}_0 = \mathbf{y}$. Also let $G$ denote the support of the current group.

### A. Implementation

LAMP initially finds a starting point or seed in every group, as in OMP. The starting point $\{j\}$ is given by $j = argmax_j |\langle \mathbf{A}_j, \mathbf{r}_n \rangle|$ where $\mathbf{A}_j$ denotes the $j$-th column of $\mathbf{A}$ and $\mathbf{r}_n$ denotes the residue after $n^{th}$ iteration. The indices are included in set $T$ as well as set $G$.

Once the starting point or seed has been found, the algorithm attempts to find the minimum block such that the entire group is contained within the block. It first searches upwards starting from $\{j\}$. The number of upward iterations is stored in $k_{up}$. It picks up the next element upwards in the vector $\mathbf{x}$ and calculates the residue including the new element in the current support of $\mathbf{x}$ i.e. $T_n$. Now, proceeding by the idea of OLS, we observe that if this element actually belongs to the support of $\mathbf{x}$ the change in the residue including this element will be significant as compared to the case where the new element does not belong to $Supp(\mathbf{x})$. We provide a theoretical justification for this statement in the next section. Thus, the new element is included in $T$ as well as $G$ only if $||\mathbf{r}_n||_2^2 - |residue(T^n \cup \{j\})|_2^2| > \varepsilon$ holds for some positive value of threshold. The iteration index $n$, upward iteration index $k_{up}$ and residue $\mathbf{r}_n$ are updated. The upward search stops if it comes across an element for which $||\mathbf{r}_n||_2^2 - |residue(T^n \cup \{j\})|_2^2| \leq \varepsilon$ since this implies that the chosen new element does not decrease the residue significantly.

---

PSEUDO CODE

**INPUT:** $A, y, K$
**OUTPUT:** $T = Supp(\mathbf{x})$
**STARTING CONDITIONS:** $T = \phi, n = 0, \mathbf{r_0} = \mathbf{y}$

---

*Repeat Until (|T|<K)*    //K-sparsity of vector x
{
　　$\{j\}$ = arg max$_j$ |<$A_j$, $r_n$>|    // find starting point in a group
　　Set n=n+1    //increase iteration index
　　$T_n=T_{n-1} \cup \{j\}$    //include this element in the support
　　$r_n$ = findresidue(Y, A, $r_{n-1}$, {j},$T_n$)
　　　　　　//find residue including this element in support
　　G=G ∪ {j}    //update the current group support

/*********Search upwards********/
　　Set $k_{up}$=1    //set the next upward iteration index
　　residue = findresidue(Y, A, $r_n$,{j-$k_{up}$}, $T_n$)
　　　　　　//find residue including the next upward element in support
　　*Repeat Until ( | |$r_n$|$^2_2$ - |residue|$^2_2$| >ε)*
　　　　　　//enter this loop only if change in residue with new element included in support is significant
　　{ Set n=n+1    //increase iteration index
　　　$T_n=T_{n-1} \cup \{j-k_{up}\}$  //include this element in the support
　　　R(n)=residue    //update residue
　　　G=G ∪ {j-$k_{up}$}    //update the current group support
　　　Set $k_{up}$=$k_{up}$+1    //update the next upward iteration index
　　　residue = findresidue (Y, A, $r_n$,{j-$k_{up}$},$T_n$)
　　　　　　//find residue after including next upward element in support
　　}
/***********Search downwards**********/
　　Set $k_{down}$=1    //set the next downward iteration index
　　residue = findresidue (y, $r_n$, {j+$k_{down}$}, $T_n$)
　　　　　　//find residue including the next downward element in support
　　Repeat Search using a similar procedure as Upward

/*******End of Group Search***********/
　　Set $G = \phi$ //Current group has been added, so G is cleared for next group search
}    // end of outer loop

Fig. 6: Pseudo Code for the proposed LAMP Algorithm

Note that a trivial condition needs to be added to the algorithm stating that the upward search automatically stops if the upward edge of the vector $\mathbf{x}$ has been reached since there would be no more rows available upward to search in that column. A similar search technique has been used for downward search too starting from $\{j\}$. The rows of $G$ finally, constitute a block of size $(k_{up} + k_{down} - 1)$. Now, the current group support $G$ is cleared, assuming the current group has been found. The algorithm returns to the outer loop and checks for the stopping condition. The search for another group continues if the total number of elements of *T* found is less than the sparsity *K*, otherwise the loop exits. The algorithm returns $T = Supp(\mathbf{x})$. A simple least squares gives the value of reconstructed **x** from $Supp(\mathbf{x})$

| PSEUDO CODE FOR *findresidue* |
|---|
| **INPUT:** $A, y, r_n, \{indices\}, T_n$ |
| **OUTPUT:** *Residue r'* |
| $S = T_n \cup \{indices\}$ <br> $r' = P_S^\perp y = (I - A\|_S (A\|_S)^\dagger) y$ |

Fig.7: The *findresidue* function as used in the algorithm in Fig. 6

LAMP collectively picks up the entire block instead of picking up the elements individually as in conventional OMP and OLS. This obviously leads to an improvement in the time complexity compared to OMP or OLS. The algorithm also has a provision for a variable group size compared to Block OMP. Moreover after the first block has been found, the next block in the group is searched just next to it. Compared with BOMP, where the location of the previously found block has no implication in the search of the other blocks in the group and all the remaining possible blocks in the vector are searched, this algorithm significantly reduces the search space for the next block in a group.

### B. Recovery Guarantee for LAMP

We prove that for a single measurement vector, LAMP exactly recovers an unknown vector **x**, of length *N* and sparsity *K* from a measurement vector **y** of dimension *M*, if the matrix A satisfies the Restricted Isometry Property (RIP) of order *K+1* with Isometry constant $\delta < 1/3\sqrt{K}$. The proof follows from Lemma 3.3 stated in the paper of Davenport Walkin on the analysis of OMP using Restricted Isometry Property [18].

We state *Lemma 3.3* and its *Corollary* from [18].

*Lemma*: *Let x' be a vector such that* $Support(x') \cap T = \phi$. *Now if A satisfies RIP of order* $|x'|+|T|+1$ *with RIP constant δ, then*

$$|\mathbf{h}_T(j) - \mathbf{x}'(j)| < \frac{\delta}{1-\delta} |\mathbf{x}'|_2 \quad \forall \quad j \notin T \quad (13)$$

*Corollary: If* $x', h, T, A$ *follow the conditions as stated in the Lemma and* $|x'|_\infty > \frac{2\delta}{1-\delta} |x|_2$, *then we are guaranteed that*

$$\arg\max_j \mathbf{h}_T(j) \in Supp(\mathbf{x}') \quad (14)$$

Assume that *T* denote the set of indices belonging to $Supp(\mathbf{x})$ obtained at each iteration and we are required to find $Supp(\mathbf{x}\|_{T^c})$. Note that, for an *N* x 1 vector **x**, an element of $Supp(\mathbf{x})$ can be picked up either as a starting point of a group, or during a search in upward or downward direction, after a starting point has been found. We have to prove that in both of these cases it is possible to pick up elements exclusively from the $Supp(\mathbf{x})$. Let us consider the two cases separately.

**Case 1: When chosen as a starting point of a group**

We proceed to prove by induction that the entering element always belongs to $Supp(\mathbf{x})$. For an *N* x 1 vector, an element $\{j\}$ is chosen as the starting point of a group using the relation $j = argmax_j < \mathbf{A}_j, \mathbf{r}_n >$. Now it has already been proven that $< \mathbf{A}_j, \mathbf{r}_n > = \mathbf{h}_{T^n}(j) \quad \forall j = 1, 2..N$ in (7) for every iteration of OMP. Thus,

$$argmax_j < \mathbf{A}_j, \mathbf{r}_n > = argmax_j \mathbf{h}_{T^n}(j). \quad (15)$$

For the first iteration i.e. at *n=0*, we have current set of indices, $T = \emptyset$ and $r_0 = y$. Now, it is to be noted that A satisfies RIP of order *K+1* with RIP constant $\delta < 1/3\sqrt{K}$. Moreover, following [18] it can be shown that,

$$|\mathbf{x}|_\infty > \frac{|\mathbf{x}|_2}{\sqrt{K}} > \frac{2\delta}{1-\delta} |\mathbf{x}|_2 \quad (16)$$

All the conditions of the *Corollary* are thus satisfied. Therefore,

$$\arg\max_j < \mathbf{A}_j, \mathbf{r}_0 > = \arg\max_j \mathbf{h}_{T^0}(j) \in Supp(\mathbf{x}) \quad (17)$$

Now by induction hypothesis, we may assume that upto the *n* th iteration, all the elements have been chosen correctly either by starting point search or upward/ downward traversal. Now, for the *n+1* th iteration, we have

$$|\mathbf{x}|_{T^c}|_\infty > \frac{|\mathbf{x}|_{T^c}|_2}{\sqrt{K-n}} > \frac{|\mathbf{x}|_{T^c}|_2}{\sqrt{K}} > \frac{2\delta}{1-\delta} |\mathbf{x}|_{T^c}|_2 \quad (18)$$

This combined with the RIP constraints on A satisfies all the conditions of the *Corollary*. This leads to the proof that for the *n+1* th iteration, we have $\arg\max_j \mathbf{h}_{T^n}(j) \in Supp(\mathbf{x}\|_{T^c})$.

**Case 2: During Upward/ Downward Search**

During the upward or downward search, we pick up an element $\{j\}$, adjacent to previous entering index if we have $\|\mathbf{r}_n\|_2^2 - |residue(T^n \cup \{j\})\|_2^2 > \varepsilon$ for some positive value of $\varepsilon$. Here $residue(T^n \cup \{j\}) = \mathbf{P}^\perp_{T^n \cup \{j\}} \mathbf{y}$ It is the residue that would be left if $\{j\}$ is included in the current support. Observe that,

$$\mathbf{r}_n = residue(T^n) = \mathbf{P}^\perp_{T^n} \mathbf{y} . \quad (19)$$

This approach is similar to OLS where at every step, $\arg\min_j |residue(T^n \cup \{j\})|$ is picked up. Following [13], minimising the function $|residue(T^n \cup \{j\})|$ is equivalent to maximising $\|\mathbf{r}_n\|_2^2 - |residue(T^n \cup \{j\})\|_2^2$. However, instead of searching the entire $T^c$ for the next index *j* that maximises $\|\mathbf{r}_n\|_2^2 - |residue(T^n \cup \{j\})\|_2^2$, we only test if the element adjacent to the previous entering index decreases the residue significantly based on a threshold $\varepsilon$. This also follows from the intuition that for group sparse signals, if one element belongs to $Supp(\mathbf{x})$ there is high probability that the adjacent element would also belong to $Supp(\mathbf{x})$ and hence would decrease the residue significantly. It thus remains to be proved that such a threshold $\varepsilon$ exists such that the function $\|\mathbf{r}_n\|_2^2 - |residue(T^n \cup \{j\})|_2^2 > \varepsilon$ only if $j \in Supp(\mathbf{x}|_{T^c})$ and is always less than or equal to $\varepsilon$ if $j \notin Supp(\mathbf{x}|_{T^c})$. If such a threshold can be determined, then it can be argued that at every upward or downward search, the algorithm picks up an element only if $j \in Supp(\mathbf{x}|_{T^c})$ and no wrong element is picked up.

Now from [13], where the authors have provided a comparison between OLS and OMP, we have the following relation.

$$|residue(T^n)|_2^2 - |residue(T^n \cup \{j\})|_2^2 = |<residue(T^n), \mathbf{b}_j>|^2 \quad (20)$$

where $\mathbf{b}_j = \begin{cases} \mathbf{a}_j/|\mathbf{a}_j|_2, & |\mathbf{a}_j|_2 \neq 0 \\ 0, & otherwise \end{cases}$

and $\mathbf{a}_j$ is the $j$th column of $\mathbf{P}_{T^n}^\perp \mathbf{A}$. Readers are referred to [35,36] for derivation. Note that since the columns of $A$ are usually normalised, $|\mathbf{a}_j|_2 \approx 1$ for $j \notin T^n$ and thus

$$\begin{aligned}|residue(T^n)|_2^2 - |residue(T^n \cup \{j\})|_2^2 &= |<residue(T^n), \mathbf{b}_j>|^2 \\ &\approx |<residue(T^n), \mathbf{a}_j>|^2\end{aligned} \quad (21)$$

It may be noted that $residue(T^n) = \mathbf{r}_n$, as defined previously.

Now, from the definition of $\mathbf{h}_{T^n} = (\mathbf{P}_{T^n}^\perp \mathbf{A})^t (\mathbf{P}_{T^n}^\perp \mathbf{A})\mathbf{x}|_{(T^n)^c}$ it follows,

$$\begin{aligned}<residue(T^n), \mathbf{a}_j> &= <\mathbf{r}_n, \mathbf{a}_j> \\ &= \mathbf{h}_{T^n}(j) \quad \forall j=1,2...N\end{aligned} \quad (22)$$

Now, we have A satisfying RIP of order $K+1$ with RIP constant $\delta$. Observe that $Supp(\mathbf{x}|_{T^c}) \cap T = \phi$ and $|x|_{T^c}| + |T| + 1 \leq K - |T| + |T| + 1 = K + 1$.

Following the conditions of the **Lemma**, we thus have

$$|\mathbf{h}_T(j) - \mathbf{x}|_{T^c}(j)| < \frac{\delta}{1-\delta}|\mathbf{x}|_{T^c}|_2 \quad \forall j \notin T \quad (23)$$

Now, for $j \notin Supp(\mathbf{x}|_{T^c})$, we have $\mathbf{x}|_{T^c}(j) = 0$.
Thus, (23) reduces to

$$|\mathbf{h}_T(j)| < \frac{\delta}{1-\delta}|\mathbf{x}|_{T^c}|_2 < \frac{\delta}{1-\delta}|\mathbf{x}|_2 \ \forall j \in T^c \text{ but } \notin Supp(\mathbf{x}|_{T^c}) \quad (24)$$

Using (24) we may thus write,

$$\begin{aligned}|<\mathbf{r}_n, \mathbf{a}_j>|^2 &= |\mathbf{h}_T(j)|^2 \\ &< (\frac{\delta}{1-\delta}|\mathbf{x}|_{T^c}|_2)^2 < (\frac{\delta}{1-\delta}|\mathbf{x}|_2)^2 \quad (25)\end{aligned}$$

$\forall j \in T^c$ but $\notin Supp(\mathbf{x}|_{T^c})$

Now, if the threshold $\varepsilon$ is chosen to be $(\frac{\delta}{1-\delta}|\mathbf{x}|_{T^c}|_2)^2$ or $(\frac{\delta}{1-\delta}|\mathbf{x}|_2)^2$ we are guaranteed that the algorithm will never pick up an element in upward or downward search if it does not belong to the support of $\mathbf{x}|_{T^c}$. This completes the proof.

Note that, we do not claim that the entire group will be picked up in the same upward or downward search if one starting point is picked up from that group. However, a significant number of elements will be picked up from the same group as long as the stopping condition is satisfied which removes the need of greedy search for every element. In general, it can be shown that as long as the adjacent element satisfies $|\mathbf{x}|_{T^c}(j)| > \frac{2\delta}{1-\delta}|\mathbf{x}|_{T^c}|_2$, the algorithm is guaranteed to pick up $\{j\}$ in the same upward or downward search. This works well for group sparse signals encountered in RADAR Imaging scenario, where the maximum values of $\mathbf{x}$ occur in a cluster (Gaussian Pulses and their derivatives).

*C. Comparison with OMP and BOMP*

The performance of LAMP is compared with OMP and BOMP to establish its utility. In this section, we have specifically considered a group sparse vector $\mathbf{x}$ of dimension 1 x 400. The vector consists of a Gaussian monocycle, covering 50 indices and is zero otherwise. Thus, sparsity of the vector (i.e. $K$) is 50 and all the non-zero elements occur together in a cluster.

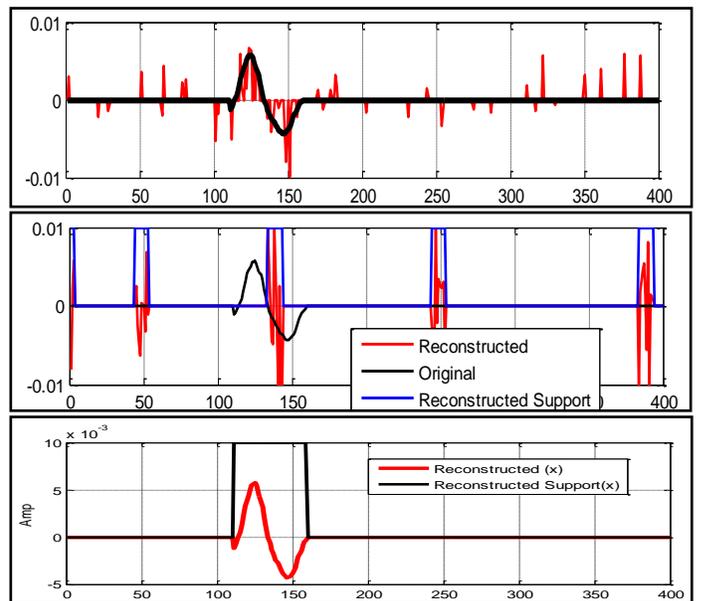

Fig 8. An Instance of failure of OMP and BOMP vs Success of LAMP (Top-OMP, Middle-BOMP, Bottom-LAMP)

Observe that Fig.8.shows an instance when OMP and BOMP( with block size 10) fails to reconstruct the original signal. Here $M$=200 measurements i.e. 50% of the original data.

Now, we proceed to simulate the exact recovery rate of LAMP as compared to OMP. It has been mentioned in [16] that an unknown vector is reconstructed perfectly if its support is exactly identified. Following [16], we have varied the number of measurements ($M$) and performed 1000 simulations for each value of $M$. The number of vectors whose support has been recovered exactly are plotted in Fig 9, both for OMP and LAMP. It can be seen clearly, that LAMP gives better recovery compared to traditional OMP algorithm, particularly when the number of measurements M is less. The selection of an element in OMP relies heavily on the properties of the sensing matrix A. However, LAMP correctly identifies a group if the first element is selected correctly. Once the starting point is found correctly, the other points are selected correctly in its vicinity. The simulations establish the utility of LAMP compared to OMP.

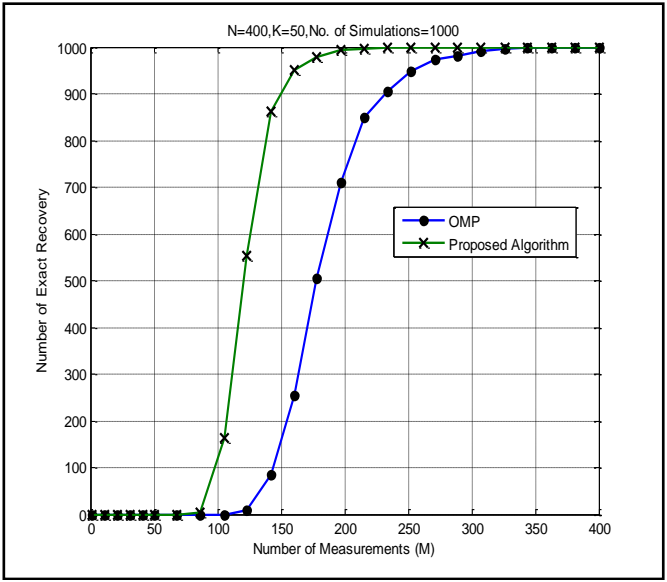

Fig 9: Exact Recovery Simulations for OMP vs LAMP

For a better comparison of relative performance of different algorithms, we introduce the notion of a relative recovery. We define the relative recovery (RR) of an algorithm as

$$RR = \frac{Cardinality(Supp(\mathbf{x}) \cap T)}{Cardinality(Supp(\mathbf{x}))} \quad (26)$$

Note that, if $Supp(\mathbf{x}) \in T$, we have $RR = 1$. In the next simulation, we have provided a Relative Recovery (RR) diagram of LAMP compared to OMP. The diagram shows a coloured plot where the index of the colour for every point $(M,S)$ is given by,

$$F(M,S) = f(\%) \quad (27)$$

if for a particular value of $M$, the reconstructed vector showed a $RR$ of atleast $f$%, for $S$ simulations out of 1000. Observe that, if $F(M,S') = f = 100\%$ for some value of $S' \in \{1,2,...,1000\}$ we have,

$$F(M,S) = 100\% \quad \forall \quad S < S' \quad (28)$$

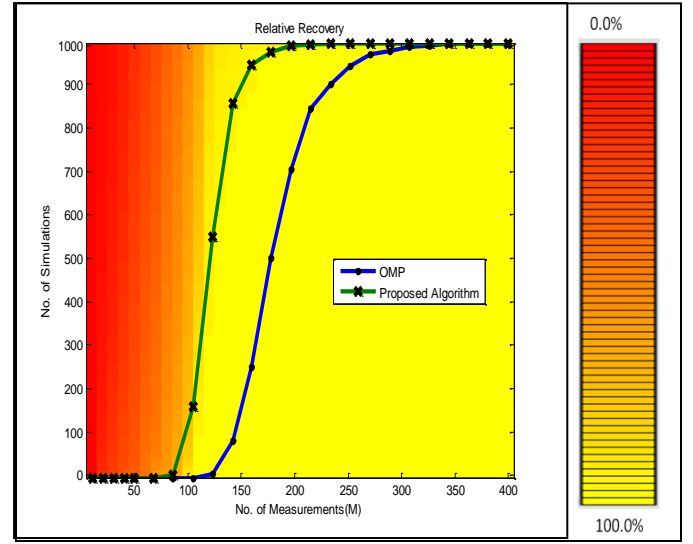

Fig 10: Relative Recovery Diagram for LAMP

Now, we compare LAMP with BOMP. Once again, we consider the same group sparse signal as described above. Note that, for simulations with various block sizes we assume that a vector $\mathbf{x}$ is correctly identified if $Supp(\mathbf{x}) \subseteq T$. Thus, $RR$ denotes the fraction of the actual Support of $\mathbf{x}$ identified correctly in the returned Support $T$. Now considering that in a practical scenario, we do not have an idea of the possible location of the clusters in the signal $\mathbf{x}$, it is incorrect to assume that the clusters can only occur at the assumed block locations in BOMP. We have thus added a variable delay with the original signal vector $\mathbf{x}$ and simulated the relative recovery($RR$) of LAMP as compared to BOMP considering the same block size 10.

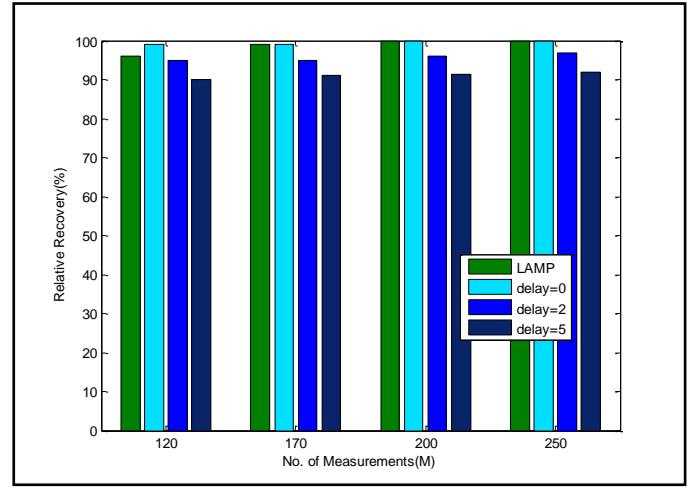

Fig11. Relative Recovery comparison of BOMP (with block size = 10) and the proposed LAMP. The return is observed at variable time-delays.

Fig.11 shows that even for the same block size 10, the performance of BOMP varies with different locations of the signal. The performance of BOMP depends on the relative location of the clusters and their extent of overlap with the possible block locations. Compared with BOMP, LAMP is well suited for identifying randomly located groups since it adaptively increases the block size, once a starting point is located.

Now we establish the flexibility of our algorithm compared to BOMP with variable block sizes. We have already demonstrated that the solution of BOMP varies with location, even for the same block size and hence, it is incorrect to assume that the clusters of **x** always overlap with the assumed block locations. We have therefore added a randomly varying delay to the original signal, for each simulation. The delay is drawn from the Uniform Distribution. For BOMP, the algorithm is stopped when number of groups identified equals *ceil (K/d)* which is the maximum number of blocks of size *d* required to cover the entire group. The simulations below show the average Relative Recovery (RR) expressed as a percentage, for BOMP with varying block sizes and randomly chosen delays compared with LAMP. In each case, 1000 simulations have been performed, for every value of *M*. We have also simulated the Mean Square Error (MSE) of the reconstructed vectors since MSE is a more reliable metric for comparing the performance of different algorithms.

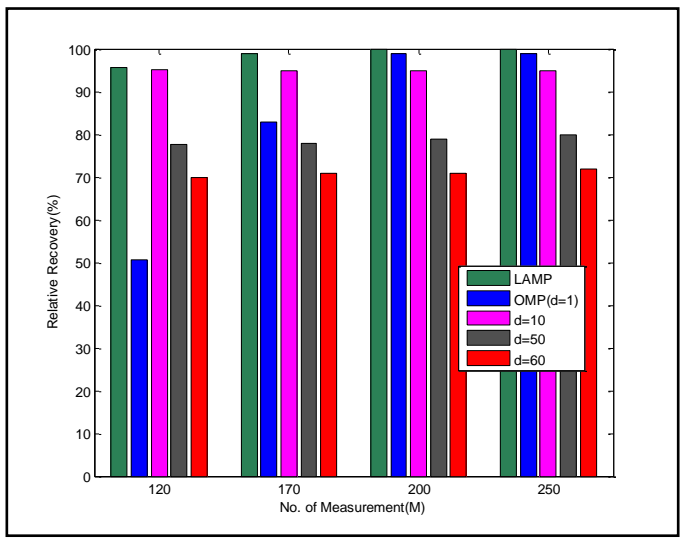

Fig.12. Comparison of Relative Recovery for BOMP for various block sizes and the proposed LAMP.

Note that, since the group length (*=sparsity K*) is 50, we have considered three blocks of sizes *d*=10, 50 and 60 respectively. A significant portion of the support is reconstructed if the blocks overlap well with the actual group location, but with randomly chosen delays and varying block sizes, the relative recovery of BOMP is usually less than LAMP.

Considering the entire range of values of *M*, LAMP is found to give the best overall performance compared to BOMP and OMP as evident from the MSE simulations in Fig.13. BOMP relies heavily on the optimal size of the blocks chosen. The algorithm thus works well only when the block size and location overlap well with the actual support of the unknown vector. The performance heavily degrades if the unknown vector has randomly located groups with variable sizes. LAMP is flexible as far as group size is concerned and does not require prior knowledge of the size of the groups. It performs much better than BOMP when the group size is not chosen optimally.

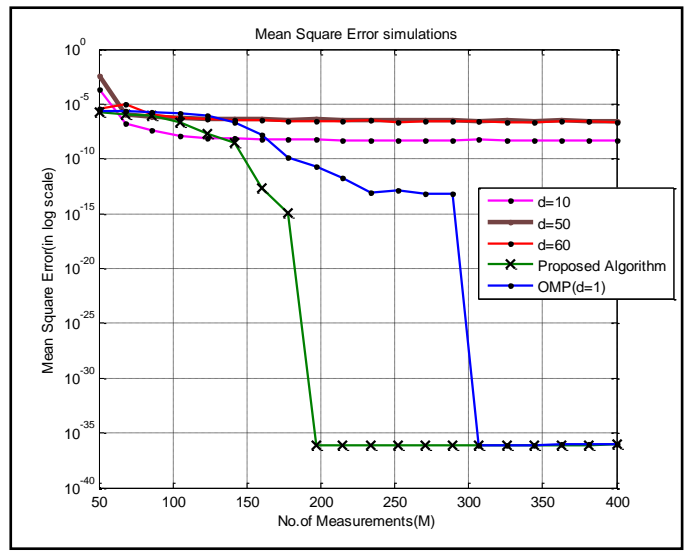

Fig. 13: Mean Square Error (MSE) comparisons of BOMP for various block sizes and the proposed LAMP.

### D. A Note On Improving Time Complexity

It has already been proved that the algorithm exactly recovers the unknown vector **x** if **A** satisfies certain conditions. Now we argue that the algorithm will recover the unknown vector x in lesser iterations compared to OMP. Clearly the time complexity of a greedy algorithm is determined by the number of times it performs the greedy search $argmax_j <\mathbf{A}_j, \mathbf{r}_n>$.

For OMP, the time complexity is proportional to $|\mathbf{x}|_0^0 = K$. However, in LAMP, observe that once a starting point in a group has been found, it picks up elements from the same group until some stopping criterion is satisfied and only then proceeds for a next greedy search. Thus, if the algorithm is stopped after a certain number of groups (*g*) has been identified, the time complexity is determined by the number of groups. Note that, the time complexity for OMP, BOMP and LAMP are *O(KMN)* [16], *O(KMN/d)* and *O(gMN)* respectively where d is the block size and *g* is the number of groups ( $g \ll K$ ).

It might also be noted that in general, the algorithm identifies the larger elements of the vector *x* first. Also, observe in UWB radar Imaging, where we mostly deal with Gaussian Pulses and its derivatives, the larger elements occur together in the same group. Thus, the algorithm in general would usually identify the larger lobes of the Gaussian Pulses together as a single group. To improve the time complexity, one might stop the algorithm once certain groups are identified, and merge the supports if they are very close vertically. This reduces the time complexity significantly compared to OMP or BOMP. However, the solution obtained on merging the supports is usually not band-limited and smooth, as compared to the original signal. In order to smoothen the final signal and restrict it to a fixed frequency band, a filtering might be performed on the least square solution obtained. This would result in a smoothened final image.

In Fig.14, the original signal and the reconstructed supports have been shown, with different stopping criterion. The signals have been reconstructed after an approximate support is found, merged together and filtered, as has been described above.

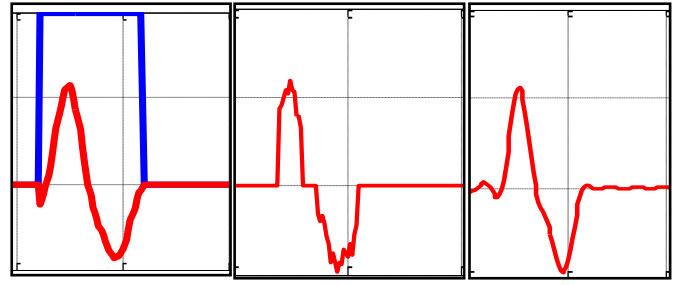

Fig16. (Left) Original Signal with Support, (Middle) Least Square solution after support estimation using LAMP, (Right) Filtered smooth signal with MSE of 7.7086e-008 before filtering and 1.0372e-036 after filtering.

The time complexity decreases as the number of greedy searches decrease. When stopped at 3 or 2 groups, the algorithm only required 3 or 2 greedy searches compared to OMP which would have required 50 such searches. We now extend the LAMP framework for reconstruction of 2D B-scan compressed data.

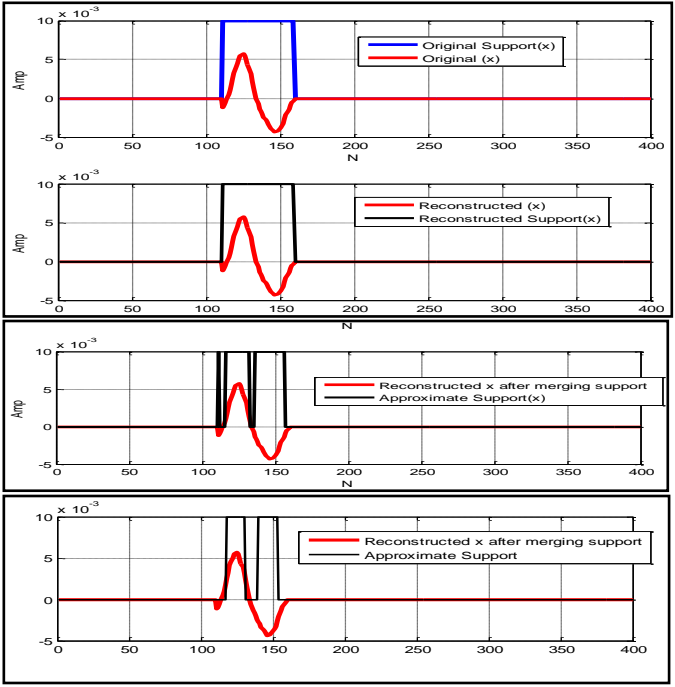

Fig 14. Signal reconstruction using LAMP with different stopping conditions (From Top to Bottom in order-Original signal, Stopped when K elements found, Stopped at 3 groups and 2 groups)

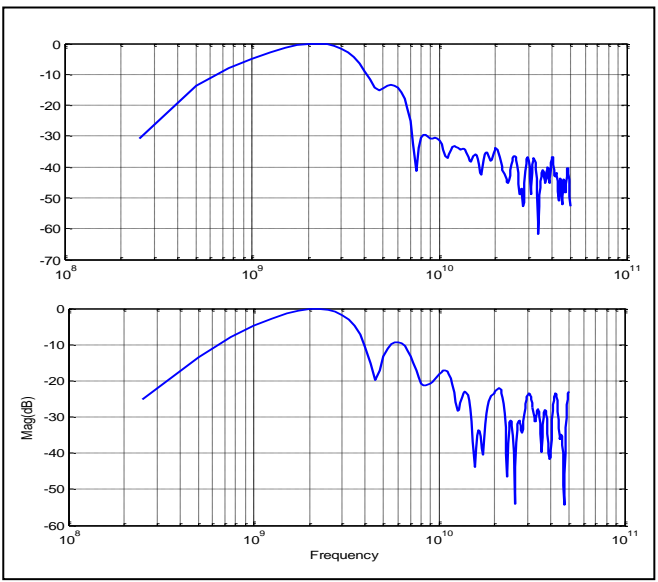

Fig15. Spectrum of Original signal (Top) and the Least Square reconstructed solution (Bottom) before filtering.

## V. EXTENSION TO MMV MODEL

In this section we propose the main extension of our algorithm in a Multiple Measurement Vector model to explore the group sparsity typically encountered in extended targets. Observe that the non-zero co-efficients in the space time data are not only clustered vertically but also horizontally and this clustering between consecutive antennae is lost if they are separately concatenated as in the SMV model. In order to preserve the clustering among neighbouring columns, we do not concatenate them into a single vector. Instead, for every B-Scan data of the buried target, we have considered a Multiple Measurement Vector (MMV) model as described below:

$$\mathbf{Y} = \mathbf{AX} \qquad (29)$$

where,

(i) $\mathbf{X} = [\mathbf{X}_1 \mathbf{X}_2 \ldots \mathbf{X}_P]$ is the original $N$ x $P$ dimensional signal matrix with each column vector $\mathbf{X}_p$ denoting the actual time domain measurements corresponding to the $p^{th}$ antennae location for $N$ samples in the time domain, where $p$ varies from 1 to $P$.

(ii) $\mathbf{A}$ is the $M$ x $N$ dimensional similar sensing matrix as described in Section III, with $M \ll N$

(iii) $\mathbf{Y} = [\mathbf{Y}_1 \mathbf{Y}_2 \ldots \mathbf{Y}_P]$ is the $M$ x $P$ matrix obtained after compressive measurements with each column denoting the compressed measurements from each antenna location. Note that the MMV model also reduces the dimension of random matrix $\mathbf{A}$ leading to ease of generation of matrix elements.

MMV model has been used in radar imaging considering signals of different polarization as stated in [37]. However, compared to the conventional Multiple Measurement Vector (MMV) model where it is assumed that the non-zero elements of Y occur in the same row, no such restrictions have been imposed in our model.

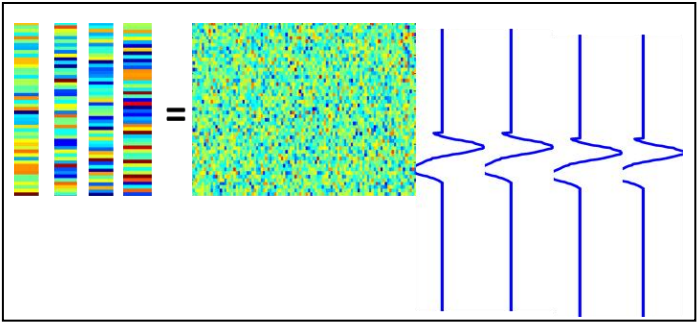

Fig 17. Schematic of Multiple Measurement Vector (MMV) Model as used in LAMP

The objective of our sparse recovery algorithm is therefore to reconstruct the matrix **X** from **Y** and **A**, with the additional information that **X** has randomly located, variably sized groups.

### A. Implementation

PSEUDO CODE

**INPUT:** $\mathbf{A}, \mathbf{Y}, K$

**OUTPUT:** $T = Supp(\mathbf{X})$

**STARTING CONDITIONS:** $T = \phi, n = 0, R_0 = Y$

*Repeat until (|T|<K)*
  { $(j,p) = \arg\max_{(j,p)} |<\mathbf{A}_j, (\mathbf{R}_n)_p>|$
    Set n=n+1      //increase iteration index
    $T_n = T_{n-1} \cup (j,p)$     //include this element in support
    $R_n$ = findresidue(Y, A ,$R_{n-1}$,[j,p] ,$T_n$)
    $G = G \cup (j,p)$
  /********Searching upward **********/
    Set $k_{up}$=1      //set the next upward index
    residue = findresidue(Y, A ,$R_n$,[j-k-,p], $T_n$)
    Repeat Until ($\| R_n \|^2_2 - | residue |^2_2 > \varepsilon$ )
    { Set n=n+1
      $T_n = T_{n-1}$ U (j-$k_{up}$ , p)
      $R_n$=residue
      G=G U (j-$k_{up}$ , p)
      Set k-=k-+1
      residue = findresidue (Y, A ,$R_n$,[j-$k_{up}$,p],$T_n$)   }
  /********Searching downward *********/
    Set $k_{down}$=1
    residue = findresidue (y ,$R_n$, [j+k+,p], $T_n$)
    Repeat Search using a similar technique as Upward
  /*******Searching left in blocks*******/
    Set $k_{left}$=1      //set left iteration index
    residue = findresidue(y ,$R_n$ , [rows(G), p-$k_{left}$] , $T_n$)
            //find residue including entire block in support
    Repeat Until ($\| R_n \|^2_2 - | residue |^2_2 > \varepsilon'$)
  //enter this loop only if change in residue with new block included in support is significant
    { Set n=n+1      //increase iteration index
      $T_n = T_{n-1}$ U [rows (G), p-$k_{left}$] //include this block in the support
      R(n)=residue
      G=G U [rows (G), p-$k_{left}$]
      Set $k_{left}$=$k_{left}$+1
      residue = findresidue (Y,A, $R_n$, [rows(G), p-$k_{left}$], $T_n$) }
  /********Searching right in blocks*****/
    Set $k_{right}$=1      //set right iteration index

    residue = findresidue (Y,A, $R_n$, [rows(G), p+$k_{right}$], $T_n$)
    Repeat search using a technique similar to left search
  /*******End of Group Search********/
    G=∅      //Current group added, so cleared for next group search
  }     // end of outer loop

Fig 18. Pseudo Code for the proposed LAMP algorithm in MMV setting.

Note that the residue is no longer a vector but a matrix denoted by $\mathbf{R}_n$. Similarly, $T \subseteq \{1,2...N\} \otimes \{1,2....P\}$ where $\otimes$ denotes the Cartesian Product of the two sets. The algorithm first identifies a particular block in a column as described previously for the SMV model. Once a set of clustered indices in a column are found, the algorithm searches in the left and right directions to find the minimum sized rectangle containing the group. The left iteration index is stored in $k_{left}$. The rows of the $p^{th}$ column that have been included in the current group search are already stored in the set $G$. These rows of $G$ constitute a block of size ($k_{up} + k_{down}$-1). Now the same rows in the next leftward column are picked up in [rows(G), b-$k_{left}$]. The residue is calculated with the new block included in the support of **X**. Intuitively, $\| \mathbf{R}_n \|^2_2 - | residue(T^n \cup \{j\}) |^2_2 > \varepsilon'$ if the new block actually belongs to the support of **X**. Thus, the block is included in the set T and the set $G$ only if $\| \mathbf{R}_n \|^2_2 - | residue(T^n \cup \{j\}) |^2_2 > \varepsilon'$ The iteration index n, the left iteration index $k_{left}$ and the residue $\mathbf{R}_n$ are updated. The left search stops if the change in residue including the next block is not significant. A similar search procedure is followed in the right direction.

Once all the four directions have been exhausted, a complete rectangular group of size $(k_{left} + k_{right} - 1)$ x $(k_{up} + k_{down} - 1)$ is found

### B. Comparison with OMP and BOMP

In this paper, we have considered a real world imaging scenario where 18 antennae are placed along a single line and two metal targets are buried at depth of 3cm and 8 cm. Corresponding to each antennae, there are 200 points in the time domain. Thus effectively, the space time data(**X**) consists of 200 x 18 pixel points.

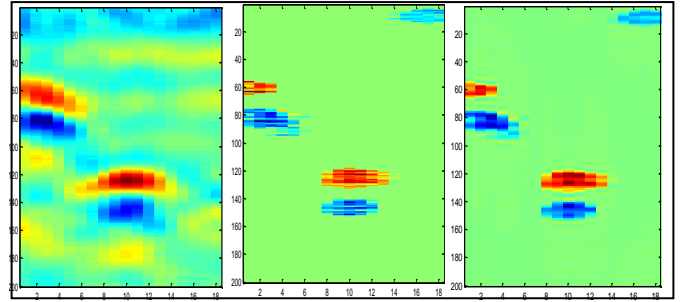

Fig 19: (Left) Original Image,(Middle) Least Square solution from merged support,(Right) Final Image after filtering

A random Gaussian Matrix **A** of dimension 40 x 200 is used to sense the data. Thus compressive measurement vector **Y** of dimension 40 x 18 is obtained. The algorithm uses only the matrix **Y** and matrix **A** as input and reconstructs the image of **X** from the compressive measurements (20%data). The results have been shown in Fig. 19.

We observe that the reconstructed images resemble the actual Image (space-time data) quite well. The major groups in the original data are in good agreement with the reconstructed images. The algorithm has been stopped once the major groups have been identified as suggested in the paper. The supports of two groups have been merged if the vertical distance between them is small. This has been found to work well for the radar Imaging scenario where the signals mainly consist of Gaussian pulses and its derivatives. The final image has been obtained after performing a filtering operation on the Least Square solution to restrict it to a fixed band-width and thereby, obtain the final smoothened image.

Now, we compare the results obtained using OMP and BOMP with that of LAMP.

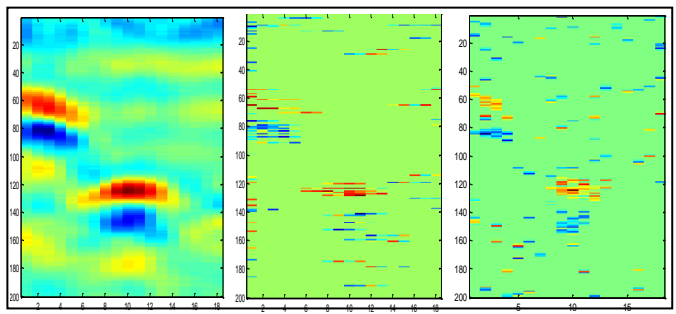

Fig 20: (Left) Original Image,(Middle) OMP reconstruction with sparsity as stopping criterion,(Right) OMP reconstruction with residue value as stopping criterion

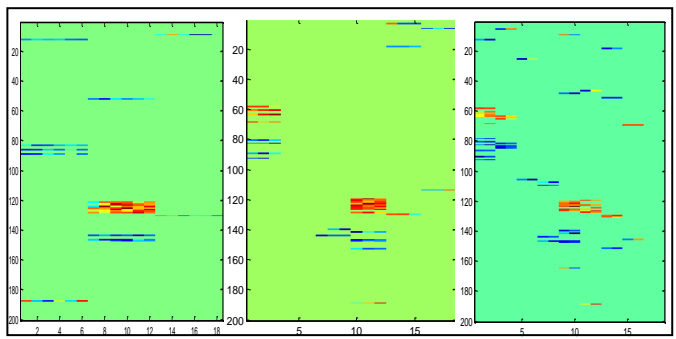

Fig. 21: BOMP reconstruction with different block sizes of 6, 3, 2 (from left in order)

Note that the images reconstructed using OMP contain significant amount of clutter and the entire group has not been identified clearly. Using BOMP, the reconstructed images vary based on the block size chosen. The results clearly establish that the BOMP algorithm is optimal only for a fixed block size, and fails to reconstruct the image accurately if the block size is not chosen correctly. The optimality of BOMP depends on prior knowledge of the best block size in which the randomly located clusters of the unknown signal would fit in.

### C. A Note on Time Complexity

The proposed framework of LAMP gives significant improvements in time complexity compared to OMP or BOMP when extended in two dimensions i.e. over a B-scan. To illustrate this, let us assume that there is a group of size, say 3 x 4. OMP requires 12 greedy searches to identify the group. BOMP with the optimal block size 3, applied in MMV as in [38] would also require 4 greedy searches for the 4 columns covered by the group. However, LAMP picks up the entire group after only one greedy search. Note that in two dimensions, the time complexity of OMP, BOMP and LAMP are $O(KMNP), O(KMNP/d)$ and $O(gMNP)$ respectively where $d$ is the block size and $g$ is the number of groups ( $g << K, K/d$ ).

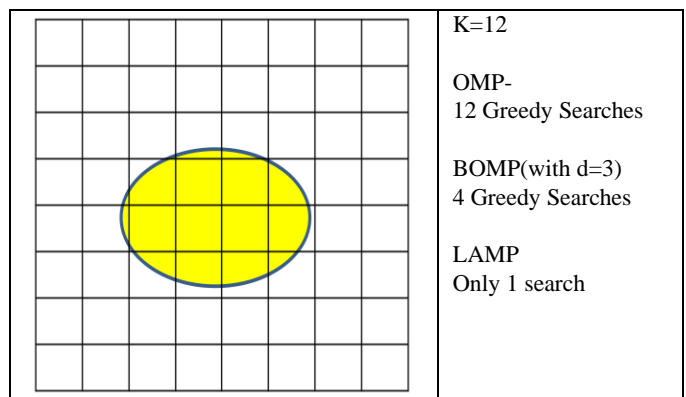

Fig. 22: An example to illustrate the gain in time complexity

K=12

OMP-
12 Greedy Searches

BOMP(with d=3)
4 Greedy Searches

LAMP
Only 1 search

## VI. DISCUSSION AND FUTURE WORK

Compressive Sensing provides a paradigm for faster acquisition of data in Ultra-wideband Radar imaging. The time domain scattered signal from the unknown buried targets appears in clusters with varying size and at arbitrary time delays. The reconstruction of such group sparse signals from compressed data using conventional techniques does not provide satisfactory results. Our proposed locally adapting matching pursuit (LAMP) provides a novel framework to address this issue. The algorithm initially locates a seed within a cluster and proceeds with identifying the non-zero elements around that locality. Following this procedure it locates all the groups in the time domain data. The algorithm is flexible and does not assume exact knowledge of the group structure. The recovery guarantee of the non-zero subsets of the time domain signal has been established. Superior performance of the algorithm in terms of the mean square error of the reconstructed signal compared to the existing techniques has been demonstrated. An extension of the LAMP framework for the B-scan data has also been proposed. The performance of the algorithm has been studied on the real-world experimental data.

LAMP offers significant advantage in terms of computational time which is roughly proportional to the number of clusters in the time domain data. Efficiency of the algorithm has been improved upon by first locating the dominant subsets of the group and then merging the supports if they are in close proximity. Further, incorporating the spectral characteristics of the transmitted pulse in the reconstruction procedure provides better accuracy in the presence of noise.

An improvement in the reconstruction can be pursued upon using total variation (TV) minimization or techniques incorporating the spectral information as a constraint while solving for the least square problem, once the support set is correctly identified using LAMP. Methods like Total Least Squares (TLS) might also be considered in future for solving such least squares problem keeping in mind of the probable error in sensing matrix. In a similar line of thought as the extension of the single measurement model to the multiple measurement model, LAMP can be further extended for adjacent B-Scans, producing the three-dimensional profile of the buried targets with enhanced accuracy and significant gain in computation time.